\documentclass[oneside]{article}
\usepackage{amsmath, enumerate, graphicx, multirow, amssymb, array, amsthm}
\usepackage[textwidth=5.5in, textheight=8in, footskip=0.45in, voffset=0.4in]{geometry}
\allowdisplaybreaks

\begin{document}
%
%********************   Page style ***************************
%
%
%********************   Title of paper **********************
%
\noindent{TechS Vidya e-Journal of Research (ISSN 2322-0791)\\ 
Vol. 2,  2013-14, pp. 24-27.}

\begin{center}
\Large\bfseries 
Maxwell's hypothesis reconsidered 
\end{center}

\begin{center}
\large\bfseries S. Satheesh
\end{center}
%
%********************   Address(s)   ************************
%
\begin{center}
Department of Applied Sciences  \\ 
Vidya Academy of Science \& Technology, Thrissur - 680 501, India.\\
e-mail:ssatheesh1963@yahoo.co.in
\end{center}
%
%********************   Abstract  ****************************
%
\begin{center}
{\bf Abstract}
\end{center}
\begin{quote}
Maxwell's derivaion of the distributions of the velocities of molecules is based on the assumption that the velocity components in the three mutualy orthogonal directions are independent. Here we note that his assumption, the phase space is isotropic, in fact nullifies the effect of a variety of dependencies among the velocity componenets. Thus we can do away with the independence assumption. Further, we observe that his conclusion regarding distribution of the velocity components (Gaussian) remains true under a set of weaker assumptions. 
\begin{center}

\end{center}
\end{quote}
%
%********************  Main text begins here ****************
%
\label{SatheeshStart}
\section{Introduction}
Maxwell derived the normal (Gaussian) model for the distribution of velocities of gas molecules based on the following assumptions (Maxwell's hypothesis, Rao, 1973, p.160).

\vspace{.2cm}
A1. The components $X, Y, Z$ of the velocity $V$ of gas molecule in the three mutually orthogonal directions are independently distributed.

A2. The marginal distributions of $X, Y, Z$ are the same.

A3. The phase space is isotropic. That is, the density of molecules with given velocity components $X, Y, Z$ is a function of the total velocity $||V|| = \sqrt(X^2 + Y^2 + Z^2)$ and not the direction.

\vspace{.2cm} 
But this assumption/ derivation of the model by Maxwell is now generally agreed to be unsatisfactory as it is based on the assumption A1 that needs to be proved first (Mayer and Mayer, 1940, p.10-12). Mayer and Mayer gave another proof of the model based on A3 and the preservation of kinetic energy under collision of molecules.\\
Here we observe that the assumption, the phase space is isotropic, in fact nullifies the effect of a variety of dependencies among the velocity components and we can still arrive at the normal model with the additional assumption of existence of second moment for $V$. Thus we can do away with the independence assumption. We also note that for certain other dependence structures the normality of $V$ is not guarenteed.

\section{Modification}

We need the following definitions in the sequel.
\vspace{.2cm} 

\noindent \textbf{Definition 2.1} A random vector $U$ in $\textbf{R}^n, n\geq 1$ integer is said to have a spherical distribution with probability density function $f(.)$ if $f(.)$ is a function only of the norm $||u||$.

\vspace{.2cm} 
If $U$ has finite second moment then its variance-covarinace matrix $\Sigma$ is of the form $\Sigma_{n \times n} = cI_{n \times n}$, where $c>0$ and $I_{n \times n}$ is the identity matrix of order $n\times n$. Thus if $U$ has a spherical distribution with finite second moment then each pair of components of $U$ are uncorrelated. See Kelker (1970) for more on spherical and elliptical distributions. \\
A note on the terminology. For the univariate normal distribution equiprobable points are equidistant from the origin and this is true for any symmetric distributions. In the case of a bivariate normal distribution, where the components are uncorrelated but have equal variance, the equiprobable contours are circles and this is true for any radially symmetric distribution. In the higher ($n$) dimensions we have: for a (multi) $n$-variate normal with variance-covariance matrix of the form $\Sigma_{n \times n} = \sigma^2I_{n \times n}$, the equiprobable contours are spheres in the Euclidean ($R^n$) space and this is true for any spherical distribution. Extending the geometry in a different direction we have: In the case of a bivariate normal distribution, where the components are uncorrelated but have unequal variance, the equiprobable contours are ellipses. In the higher dimensions we have elliptical (elliptically contoured) distributions whose equiprobable contours are ellipsoids in the $R^n$ space. For example, we have the $n$-variate normal distribution with variance-covariance matrix $\Sigma_{n \times n}$ that is positive definite.

\vspace{.2cm}  
A consequence of spherical symmetry is that if a spherically symmetric distribution has a density then it will be a function only of the norm $||X||$ of $X$. More generally $X$ has a spherical distributions if $X$ and $HX$ have the same distribution for all $n \times n$ orthogonal matrices $H$.

\vspace{.2cm}     
\noindent \textbf{Definition 2.2} (Lehman, 1966) Two random variables $X$ and $Y$ are said to be positively quadrant dependent (PQD) if for all $(x, y) \in \textbf{R}^2$
\[P\{X \leq x, Y \leq y\} \geq P\{X \leq x\} P\{Y \leq y\}.\]

\vspace{.2cm} 
Clearly, A3 implies that the distribution of the velocities of the molecules with given velocity components is spherical, that is, $V$ is spherical. If we assume that $V$ has finite second moment then each velocity component is uncorrelated with the other two. An implication of this is that there is no linear dependence between any two of the components, (Feller, 1968, p.236). It is known (Lehman, 1966) that if two random variables are PQD then non-correlation implies independence. Thus if the component pairs of $V$ are assumed to be PQD then they are independent. Since $V$ is spherical, independence of the components of $V$ is possible only if $V$ has a normal distribution, (Kelker, 1970). Thus Maxwell's hypothesis may be modified as

\vspace{.2cm} 
B1. The density of molecules with given velocity components $X, Y, Z$ in three mutually orthogonal directions, is a function of the total velocity $||V|| = \sqrt(X^2 + Y^2 + Z^2)$ and not the direction.
 
B2. The velocity vector $V=(X, Y, Z)$ has finite second moment.

B3. The component pairs of $V$ are PQD.

\vspace{.2cm}
 
In fact instead of $B3$ we need assume only that any of the component pairs of $V$ are PQD. Then the non-correlation between them implies they are independent and consequently this pair has a bivariate normal distribution. Finally, the normality of a component pair implies that the whole vector $V$ has a normal distribution. Still narrowing down; the normaility of any one of the velocity components and $B1$ is enough for $V$ to be normal, Kelker (1970).  

\section{Discussion} Some other concepts of bivariate positive dependence, to mention a few, are: association (Esary \textit{et al.}, 1967), regression dependence and likelihood ratio dependence (Lehman, 1966), which are successively stronger. Since association implies PQD we can assume any of these stronger forms of dependence in B3 and still conclude that $V$ is normal. All these concepts have got their negative analogs and we can assume any of these as well in B3  to arrive at the same conclusion.

\vspace{.2cm} 
However, one should not be led to think that we can assume any form of dependence between the pairs in $V$ and still conclude that $V$ is normal. For example, we have the class $H(2)$ of bivariate distributions (Jogdeo, 1968) with a certain dependence structure that is a generalisation of regression dependence, and $H(2)$ has no inclusive relations with the class of PQD distributions. Hence we cannot invoke the non-correlation in the component pairs of $V$ implicit in B1 and B2, and conclude that $V$ is normal.

\vspace{.2cm} 
Trivariate dependence concepts are also discussed in the literature. But they do not add very much to our understanding of the problem in the present set up for the following reasons. Some of them such as , the class $L(3)$ of trivariate distributions (Jogdeo, 1968), orthant dependence (Joag-dev, 1983) and association in trivariate distributions (Esary \textit{et al.}, 1967), to mention a few, implies PQD and thus the problem essentially reduces to that we have already discussed. Some other concepts demand more than non-correlation in the component pairs for them to be independent; \textit{e.g} we have the class $H(3)$ of trivariate distributions  (Jogdeo, 1968). But as we have mentioned earlier, the non-correlation in the component pairs is an implication of B1 and B2, and B1 is very fundamental to our problem.

\vspace{.2cm} 
The above discussion brings to light the fact that the lack of correlation (linear dependence) in the component pairs of the velocity vector nullifies the effect of a variety of dependencies in it. In the literature of statistical distribution theory this is not a new result, see \textit{e.g} Sampson (1983, Theorem 2.1), but this re-examination of the Maxwell's derivation is. However, this idea is not still well known in the literature, it seems; see \textit{e.g} Terrell (1999, p. 379). Also, our discussion suggests that there are potential areas where we can make good use of the non-correlation inherent in the assumptions, instead of assuming the stronger notion of independence, and still arrive at the conclusion.

\subsection*{References}

Esary, J D; Proschan, F and Walkup, D W (1967), Association of random variables with applications, \textit{Ann. Math Statist.}, \textbf{38}, 1466-1474.

Feller, W (1968), \textit{An Introduction to Probability Theory and Its Applications}, Vol. 1, $3^{rd}$ Edn., Wiley, New York.

Joag-Dev, K (1983), Independence via uncorrelatedness under certain dependence structures, \textit{Ann. Probab.}, \textbf{11}, 1037-1041.

Jogdeo, K (1968), Characterisations of independence in certain families of bivariate and multivariate distributions, \textit{Ann. Math. Statist.}, \textbf{39}, 433-441.

Kelker, D (1970), Distribution theory of spherical distributions and a location-scale parameter generalisation, \textit{Sankhya A}, \textbf{32}, 419-430.

Lehman, E L (1966), Some concepts of dependence, \textit{Ann. Math. Statist.}, \textbf{37}, 1137-1153.

Mayer, J E and Mayer, M G (1940), \textit{Statistical Mechanics}, Wiley, New York.

Rao, C R (1973), \textit{Linear Statistical Inference and Its Applications}, $2^{nd}$ Edn., Wiley New York.

Sampson A R (1983), Positive dependence properties of elliptically symmetric distributions, \textit{J. Multi. Anal.}, \textbf{13}, 375-381.

Terrell, G R(1999), \textit{Mathematical Statistics: A Unified Introduction}, Springer, New York.

\end{document}